\documentclass[12pt,preprint]{aastex}
\usepackage{graphicx,amsmath,amssymb}
\usepackage{pdflscape}
\usepackage{url}

\newcommand{\Order}{\ensuremath{\mathcal{O}}}

\begin{document}
\title{UKIRT microlensing surveys as a pathfinder for {\it WFIRST}:\\
The detection of five highly extinguished low-$|b|$ events}

\author{
Y. Shvartzvald\altaffilmark{1,a},
G. Bryden\altaffilmark{1},
A. Gould\altaffilmark{2,3,4},
C.~B.~Henderson\altaffilmark{1,a},
S.~B.~Howell\altaffilmark{5},
C.~Beichman\altaffilmark{6}
}
\altaffiltext{1}{Jet Propulsion Laboratory, California Institute of Technology, 4800 Oak Grove Drive, Pasadena, CA 91109, USA}
\altaffiltext{2}{Max-Planck-Institute for Astronomy, K\"onigstuhl 17, 69117 Heidelberg, Germany}
\altaffiltext{3}{Korea Astronomy and Space Science Institute, Daejon 305-348, Republic of Korea}
\altaffiltext{4}{Department of Astronomy Ohio State University, 140 W.\ 18th Ave., Columbus, OH 43210, USA}
\altaffiltext{5}{Kepler \& K2 Missions, NASA Ames Research Center, PO Box 1,M/S 244-30, Moffett Field, CA 94035}
\altaffiltext{6}{NASA Exoplanet Science Institute, California Institute of Technology, Pasadena, CA 91125, USA}
\altaffiltext{a}{NASA Postdoctoral Program Fellow}

\begin{abstract}
Optical microlensing surveys are restricted from detecting events near the Galactic plane and center,
where the event rate is thought to be the highest, due to the high optical extinction of these fields.
In the near-infrared (NIR), however, the lower extinction leads to a corresponding increase in event detections and is a primary driver for the
wavelength coverage of the {\it WFIRST} microlensing survey.
During the 2015 and 2016 bulge observing seasons we conducted NIR microlensing surveys with UKIRT in conjunction with and in support of the
{\it Spitzer} and {\it Kepler} microlensing campaigns.
Here we report on five highly extinguished ($A_H=0.81-1.97$), low-Galactic latitude ($-0.98\le b\le -0.36$) microlensing events discovered from our 2016 survey.
Four of them were monitored with an hourly cadence by optical surveys but were not reported as discoveries, likely due to the high extinction.
Our UKIRT surveys and suggested future NIR surveys enable the first measurement of the microlensing event rate in the NIR.
This wavelength regime overlaps with the bandpass of the filter in which the {\it WFIRST} microlensing survey will conduct its highest-cadence observations,
making this event rate derivation critically important for optimizing its yield. 

\end{abstract}


{\section{{Introduction}
\label{sec:intro}}

In recent years microlensing has established itself as an important technique for exoplanet detection and characterization. 
Since the first planet detected via microlensing over a decade ago \citep{Bond.2004.A}, 40 exoplanets have been discovered and published\footnote{NASA Exoplanet Archive: \url{ http://exoplanetarchive.ipac.caltech.edu/}}
in a variety of planetary systems, including triple systems with one (e.g., \citealt{Gould.2014.A}) or two (e.g., \citealt{Gaudi.2008.A}) planets.
Furthermore, microlensing is currently the only method that can systematically probe the free-floating planet population \citep{Sumi.2011.A}.
The ability of a microlensing experiment to detect these systems relies heavily on the underlying microlensing event rate,
which itself depends on the surface density of source and lens stars toward the survey fields.
Therefore, microlensing surveys have concentrated their observing efforts toward the Galactic bulge, where the stellar surface density is highest.
However, traditionally these surveys have been conducted in optical wavelengths, which suffer from high extinction near the Galactic plane and center,
and thus their fields were selected in order to balance the stellar surface density with the extinction.

Extinction decreases as wavelength increases. Therefore, near-infrared (NIR) microlensing surveys enable observations closer to
the Galactic plane and center, where the event rate is expected to be higher \citep{Gould.1995.A}.
Understanding this potential, the {\it WFIRST} flagship mission will conduct a ∼432-day NIR microlensing survey toward the Galactic
bulge divided equally between six 72-day seasons \citep{Spergel.2015arXiv.A}. This survey will revolutionize our understanding of bound planetary systems with
separations of 1--10 AU as well as free-floating exoplanets.
However, until now no dedicated NIR microlensing surveys have been conducted, and so the event rate and the microlensing optical depth in the NIR have not been measured.
VISTA Variables in the V{\'i}a L{\'a}ctea (VVV) conducted a variability survey toward the inner Milky Way in the NIR, with multiple epochs with the $K_s$ filter \citep{Minniti.2010.A}.
However, the survey cadence was irregular and too low to routinely detect microlensing events, which have a typical event timescale of $\sim$20 days.  
Currently, the VVV survey has published only one microlensing event \citep{Minniti.2015.A}, which had a long timescale and was magnified for $\sim$100 days,
and thus the VVV cadence was sufficient.
This event was in a low-extinction field and was independently detected\footnote{\url{http://ogle.astrouw.edu.pl/ogle4/ews/ews.html}} in $I$-band by the OGLE  microlensing survey \citep{Udalski.2015.B}.

NIR follow-up observations have been performed for microlensing events, usually in $H$-band.
These observations lead to a measurement of the source flux in the observed NIR filters, a necessary component for the interpretation of post-event NIR adaptive optics (AO) observations.
The AO observations resolve out stars not related to the event, isolating the microlensing target (source plus lens) and facilitating a measurement (or constraint) of the lens flux.
This can break inherent degeneracies and enable a derivation of the physical
properties of the lensing system, including its mass and distance.

In each of the last two bulge observing seasons we conducted dedicated NIR microlensing surveys with UKIRT. These observations were carried out in conjunction
with the space microlensing surveys carried with {\it Spitzer} (2015) and {\it K2} (2016). The UKIRT observations provided the NIR coverage required for future lens flux
measurements as well as the cadence necessary for event detection and characterization. UKIRT observations were crucial for source characterization and anomaly coverage in  the detection of a massive remnant in a wide binary \citep{Shvartzvald.2015.A}.
These surveys can also facilitate a preliminary estimation of the NIR event rate, although their total combined duration and area
are currently not sufficient to precisely constrain it.

In this letter we present the detection of five microlensing events discovered by the 2016 UKIRT survey, all of which are located in highly extinguished fields.
These five detections are the result of an initial search for events within the 2016 UKIRT data set.
As such, they emphasize the richness of the existing UKIRT microlensing survey data set as well as the full potential that can be realized by a long-term NIR microlensing survey toward the inner bulge.
In Section \ref{sec:obs}, we detail the observation and reduction procedures of our 2015--2016 surveys.
In Section \ref{sec:events}, we present the five new microlensing events and the extinction and reddening toward each,
and finally we discuss the implications of these detections in Section \ref{sec:discussion}.

{\section{Observational data and reduction}
\label{sec:obs}}

\subsection{Observational setup}
The observations were carried out with the wide-field NIR camera (WFCAM) at
the UKIRT 3.8m telescope on Mauna Kea, Hawaii.
The field of view of each of the four WFCAM detectors is $13.6'\times13.6'$ and the detector layout contains gaps whose areas are 94\% of one detector.
All observations were in $H$-band, with each epoch composed of sixteen 5-second co-added dithered exposures (2 co-adds, 
2 jitter points, and $2\times2$ microsteps).

\subsection{2015 northern bulge survey}

In 2015 we conducted a 120 hr, 3.4 deg$^2$ UKIRT survey in conjunction with the 2015 $Spitzer$ microlensing campaign.
Our survey focused on northern bulge fields (see Figure \ref{fig:fields}), which suffer from high extinction.
We thus complement the sparse optical coverage with higher-cadence, high signal-to-noise coverage in the NIR, enabling future lens flux measurements.
18 UKIRT fields were selected using the procedures developed by
\citet{Poleski.2016.B}, who showed that the product of the surface density
of stars with $I<20$ and clump stars is a good estimator of the microlensing
event rate.
The survey was conducted over 39 days, spanning HJD' (HJD - 2450000) = 7180--7219, with a nominal cadence of 5 observations per night.
During this period, 32 microlensing events discovered by OGLE were ongoing,
and we recover the UKIRT light curve for all of them.

\subsection{2016 southern bulge survey}

In 2016, we carried out a second UKIRT survey, observing for 240 hrs in conjunction with Campaign 9 of the $K2$ Mission ($K2$C9),
which conducted the first automated microlensing survey from space \citep{Henderson.2015arXiv.A}.
The 2016 UKIRT survey supported the $K2$C9 survey, which observed toward the southern bulge, giving NIR coverage of events inside the survey superstamp.
32 UKIRT fields were selected covering 6 deg$^2$, including the entire $K2$C9 superstamp and extending almost to the Galactic plane (see Figure \ref{fig:fields}).
The survey was conducted over 91 days, spanning HJD' = 7487--7578, with a nominal cadence of 2--3 observation per night (limited by the bulge visibility at the beginning of the season).

\subsection{Data reduction}

The UKIRT dithered images were reduced, astrometrically calibrated, and stacked by the Cambridge Astronomy Survey Unit (CASU; \citealt{Irwin.2004.A}).
Then, three different photometry methods were applied to the data in order to extract the light curves:
\begin{enumerate}
 \item Multi-aperture photometry --- standard 2MASS-calibrated photometry catalogs produced by CASU for all UKIRT/WFCAM images \citep{Hodgkin.2009.A}.
 In the crowded fields of the Galactic bulge this method works well only for bright (and/or completely isolated) stars, including partially saturated stars.
 \item PSF photometry --- using SExtractor \citep{Bertin.1996.A} and PSFEx \citep{Bertin.2011.A} and calibrated to 2MASS.
 This method works well even for crowded fields but not for saturated stars (thus making it complementary to method 1).
 \item DIA photometry --- using the pySis software \citep{Albrow.2009.A}. This method yields even better results for faint events and gives smaller scatter than the first two methods.
 However, currently only the target DIA flux is measured, without photometric alignment to a standard system.
\end{enumerate}

\subsubsection{$K2$C9 products}

UKIRT light curves for all microlensing events detected within the $K2$C9 survey superstamp, whether by ground-based surveys and/or $K2$, will be uploaded to the $K2$C9 ExoFOP site\footnote{\tt https://exofop.ipac.caltech.edu/k2/microlensing}.
Currently, only the light curves of known anomalous events (i.e., events for which the light curve deviates from that expected for a single-lens microlensing event) have been uploaded to the website.
Along with each light curve produced with methods 1 and 2, we also uploaded the photometric catalog of stars within 2' of each microlensing target.

{\section{Microlensing events}
\label{sec:events}}

\subsection{Detections}

The detection of a microlensing event is not a trivial task.
Since the microlensing optical depth toward the Galactic bulge is $\Order(10^{-6})$, millions of light curves need to be searched
in order to detect an event. We applied a new event detection algorithm (Kim et al., in preparation) that uses microlensing models calculated from a grid of times of peak 
magnification, $t_{0}$, and effective timescales, $t_{\rm eff}$,
to search for light curves characteristic of microlensing.

In this work we analyzed only 7 of the 132 subfields (each UKIRT field is composed of 4 subfields, corresponding to the 4 detectors) from the 2016 survey.
We chose the 7 subfields closest to the Galactic center, one of which is also the subfield closest to the Galactic plane.
Each subfield contains $\sim$60000 sources detected in at least 50 out of 152 epochs.
We selected events for which $t_{0}$ occurs during the UKIRT survey and that also show clear baseline.
Here we present five new microlensing events, none of which were discovered by the optical microlensing surveys, OGLE and MOA \citep{Sumi.2003.A}, as indicated by their alert pages.
Figure \ref{fig:lcs} shows the light curves of these events, which we designate UKIRT1--5.

Events UKIRT(2,3,4,5) lie in OGLE-IV field 500 and were observed by OGLE with a $\sim$1 hr cadence, which nominally would be sufficient to detect them, given their timescales are all $t_{\rm E} \gtrsim 7$ d, but for the extinction.
UKIRT1, on the other hand, resides in OGLE-IV field 533, which is observed only occasionally.

\subsection{Extinction and reddening}

We use VVV catalogs of stars within $2'$ of each event to construct $(H-K_{s},H)$ color-magnitude diagrams (CMDs; see Figure \ref{fig:cmd}).
We identify the red clump position in each CMD and derive the extinction and reddening in each field by comparing these values to the intrinsic clump values.
Specifically, the intrinsic clump color used is $(H-K_{s})_{0,cl} = 0.1$ and its $H$ magnitude slightly varies as a function of $l$, with $H_{0,cl} = 13.15$ at the Galactic center
\citep{Nataf.2013.A}.
Table \ref{tab:ev_details} gives for each event its Equatorial and Galactic coordinates, the intrinsic magnitude and position of the red clump and the derived extinction and reddening.

UKIRT4 suffers from the highest extinction, with $A_H=1.97$, while UKIRT5 has the lowest extinction, with $A_H = 0.81$.
However, the event baseline magnitude for UKIRT5 is sufficiently faint ($H=16.3$) that it was not detected in optical.
There are no publicly available deep optical catalogs of these fields that can be used to detect the red clump. Thus, the optical extinction cannot be measured directly.
While a typical extinction ratio in the Galaxy is $A_I/A_H\sim3.5$, this ratio can increase significantly for high extinctions
and so sets only a lower limit on the optical extinction. 

\subsection{Models}

A standard single-lens microlensing model with three ``Paczy{\'n}ski'' \citep{Paczynski.1986.A} parameters, $(t_0,t_{\rm E},u_0)$, was fitted to each light curve,
and two additional flux parameters accounting for the source flux, $f_s$ (which can be converted to the $H$-band source magnitude, $H_s$), and possible blend flux, $f_b$.
As is common for single-lens events, we find a strong correlation between $t_{\rm E},u_0$, and the blending fraction, defined as $f_{bl}=f_s/(f_s+f_b)$
(in this convention $f_{bl}=1$ means no blending).  
Table \ref{tab:ev_pars} gives the best-fit model for each event.
Since we had observations only with $H$-band, we cannot derive the source color and thus its type.
Nevertheless, assuming the source is a bulge star, we estimate the source type based on its magnitude, $H_s$, relative to that of the red clump.
This analysis suggests that the source of UKIRT1 is a clump star, the sources of UKIRT2--3 are red giant branch stars, and the sources of UKIRT4--5 are turnoff stars.
 
{\section{Discussion}
\label{sec:discussion}}

We present the first microlensing events detected from our 2016 UKIRT survey.
This, along with our 2015 UKIRT survey, constitutes the first dedicated NIR microlensing program.
The events were detected in highly extinguished fields and thus were not detected by current optical microlensing surveys.
These events are located close to the Galactic plane and center and give a preliminary indication of the NIR event rate in these fields,
which should be higher than that in the optical fields.
The 2015--2016 UKIRT fields are mostly overlapping with the optical microlensing surveys and thus will allow us, after analysis of our complete dataset,
to compare the number of events detected with UKIRT between different fields and give preliminary estimates of the relative event rate.

The {\it WFIRST} microlensing survey will primarily use a wide NIR filter in the range of 0.927--2.0$\mu$m \citep{Spergel.2015arXiv.A}.
Our $H$-band UKIRT microlensing surveys are the first surveys conducted in a bandpass that overlaps with the highest-cadence filter of the {\it WFIRST} microlensing survey.
The sources in our sample represent only the bright end of the source luminosity function and thus suggest that the NIR event rate in these fields
will be high for {\it WFIRST}, given its expected photometric depth.
All five events are at the edges of the currently proposed {\it WFIRST} fields (see Figure \ref{fig:fields})
and thus perhaps indicate that these fields should be moved even closer to the Galactic center.

Over 100 events detected by optical surveys and observed by $Spitzer$ and $K2$C9 were ongoing during our UKIRT campaigns.
For the vast majority of these, UKIRT observations will give $H_{s}$.
This will allow for systematic AO follow-up observations of these targets in order to measure their lens fluxes and derive their physical properties.
The lens flux method will be the primary channel for determining the physical properties of {\it WFIRST} microlensing planets \citep{Yee.2015.B}.
The UKIRT sample can be thus used to help validate this technique using events with full characterization via the microlens parallax measurements from $Spitzer$ and $Kepler$.

The relatively short duration of our experiment (39 and 91 days in 2015 and 2016, respectively) limited the number and types of events we can detect.
While searching for events in this preliminary exploration of the UKIRT data, we found many light curves 
potentially indicative of microlensing events but for which the observations covered only the rising or falling portion of the light curve.
In other cases, only the peak was observed, with no baseline data (suggestive of a long-timescale event, if indeed microlensing).
Understanding these limitations and their implications for the detection efficiency of planets is important in the context of {\it WFIRST},
since each of its microlensing campaigns is planned to be 72 days long. Some of the event properties can be constrained using data from other seasons.
In particular, the validation that it is truly a microlensing event (thus appears constant in other seasons).

\acknowledgments
Work by YS and CBH was supported by an
appointment to the NASA Postdoctoral Program at the Jet
Propulsion Laboratory, administered by Universities Space Research Association
through a contract with NASA.
Work by AG was supported by NSF grant AST-1516842.
The United Kingdom Infrared Telescope (UKIRT) is supported by NASA and
operated under an agreement among the University of Hawaii, the University
of Arizona, and Lockheed Martin Advanced Technology Center; operations are
enabled through the cooperation of the Joint Astronomy Centre of the Science
and Technology Facilities Council of the U.K.
We acknowledge the support from NASA HQ for the UKIRT observations in connection with $K2$C9.
Based on data products from observations made with ESO Telescopes at the La Silla Paranal Observatory
under programme ID 179.B-2002.

Copyright 2016. All rights reserved.


\begin{table}
\centering
\caption{Event coordinates, estimated red clump intrinsic magnitude and position, extinction, and reddening.
\label{tab:ev_details}}
\begin{tabular}{c|cccc|ccccc}
\tableline\tableline
\# & RA	& Dec & $l$ & $b$ & $H_{0,cl}$ & $(H-K_s)_{cl}$ & $H_{cl}$ & $A_H$ & $E(H-K_s)$	\\
 & [J2000]	& [J2000] & [deg] & [deg] & [mag] & [mag] & [mag] & [mag] & [mag]\\
\tableline\\[-10pt]
1 & 17:50:40.9 & -27:47:10.30 & 1.56 & -0.36 & 13.09 & 1.07 & 15.00 & 1.91 & 0.97\\
2 & 17:50:26.5 & -28:02:44.32 & 1.31 & -0.45 & 13.10 & 0.83 & 14.65 & 1.55 & 0.73\\
3 & 17:50:50.9 & -28:16:54.52 & 1.15 & -0.64 & 13.11 & 0.62 & 14.10 & 0.99 & 0.52\\
4 & 17:50:24.6 & -28:50:33.32 & 0.62 & -0.85 & 13.13 & 1.10 & 15.10 & 1.97 & 1.00\\
5 & 17:51:05.3 & -28:51:24.69 & 0.69 & -0.98 & 13.12 & 0.55 & 13.93 & 0.81 & 0.45\\
\tableline\tableline
\end{tabular}
\end{table}

\begin{table}
\centering
\caption{Best-fit microlensing model parameters
\label{tab:ev_pars}}
\begin{tabular}{c|ccc|ccc}
\tableline\tableline
\# & $t_0$ & $u_0$ & $t_{\rm E}$ & $H_s$ & $f_{bl}$ & Source \\
  & [HJD'] &  & [days] & [mag] &  & type \\
\tableline\\[-10pt]
1 & 7549.16  & 0.32 &  7.4 & 14.9 & 1.00 & clump\\
2 & 7502.86  & 0.18 & 12.2 & 16.3 & 0.27 & red giant branch\\
3 & 7507.70  & 0.24 & 28.5 & 15.9 & 0.27 & red giant branch\\
4 & 7497.68  & 0.16 & 29.5 & 18.6 & 0.04 & turnoff\\
5 & 7548.32  & 0.31 &  8.5 & 17.6 & 0.31 & turnoff\\
\tableline\tableline
\end{tabular}
\end{table}

\begin{figure}
\plotone{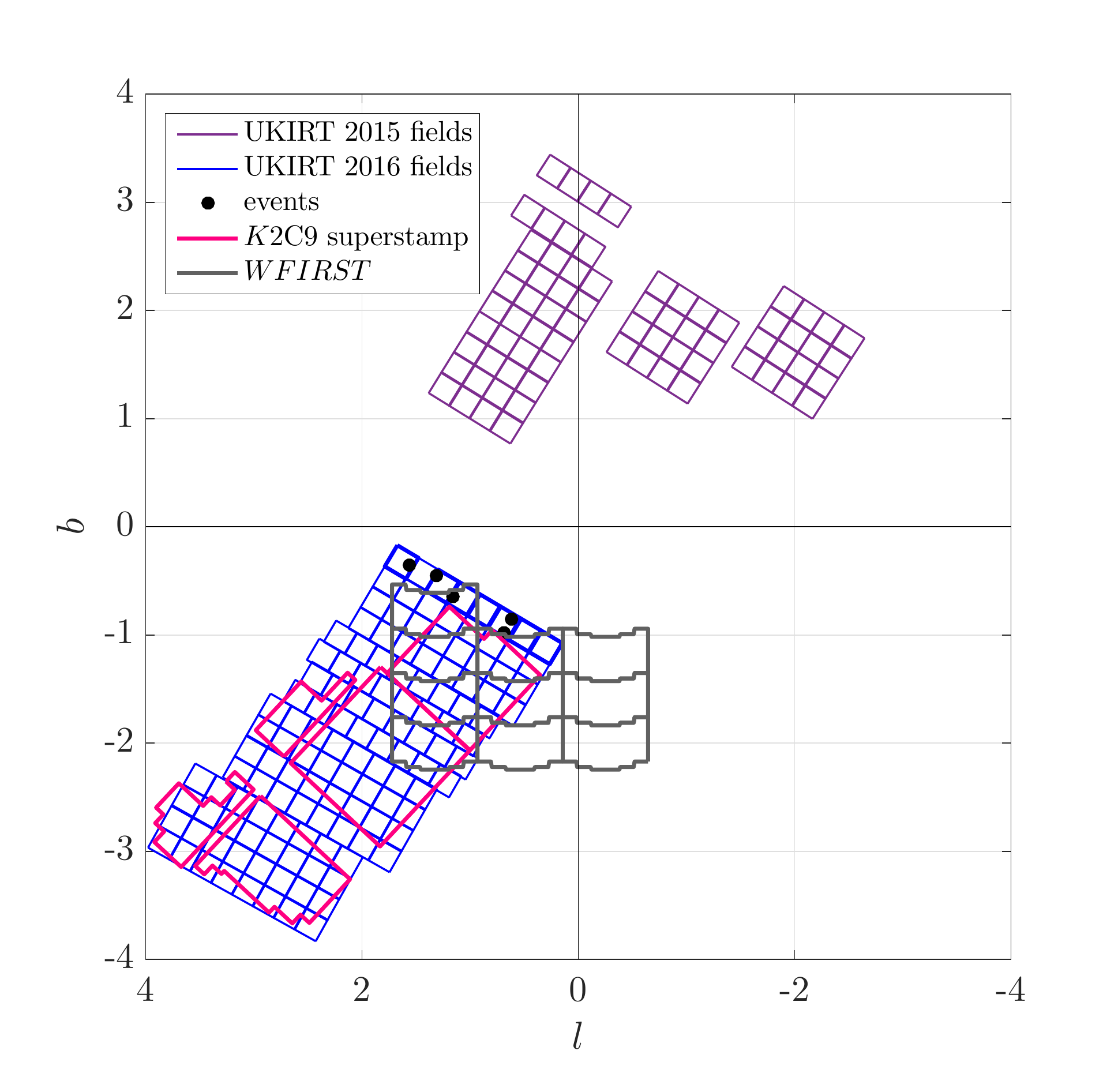}
\caption{UKIRT 2015 (purple) and 2016 (blue) microlensing survey fields.
Fields with thick lines are the seven 2016 fields searched in this work. 
The black circles are the positions of the detected events.
Also shown are the $K2$C9 superstamp (pink) and the proposed {\it WFIRST} fields (gray).
}
\label{fig:fields}
\end{figure}

\begin{figure}
\plotone{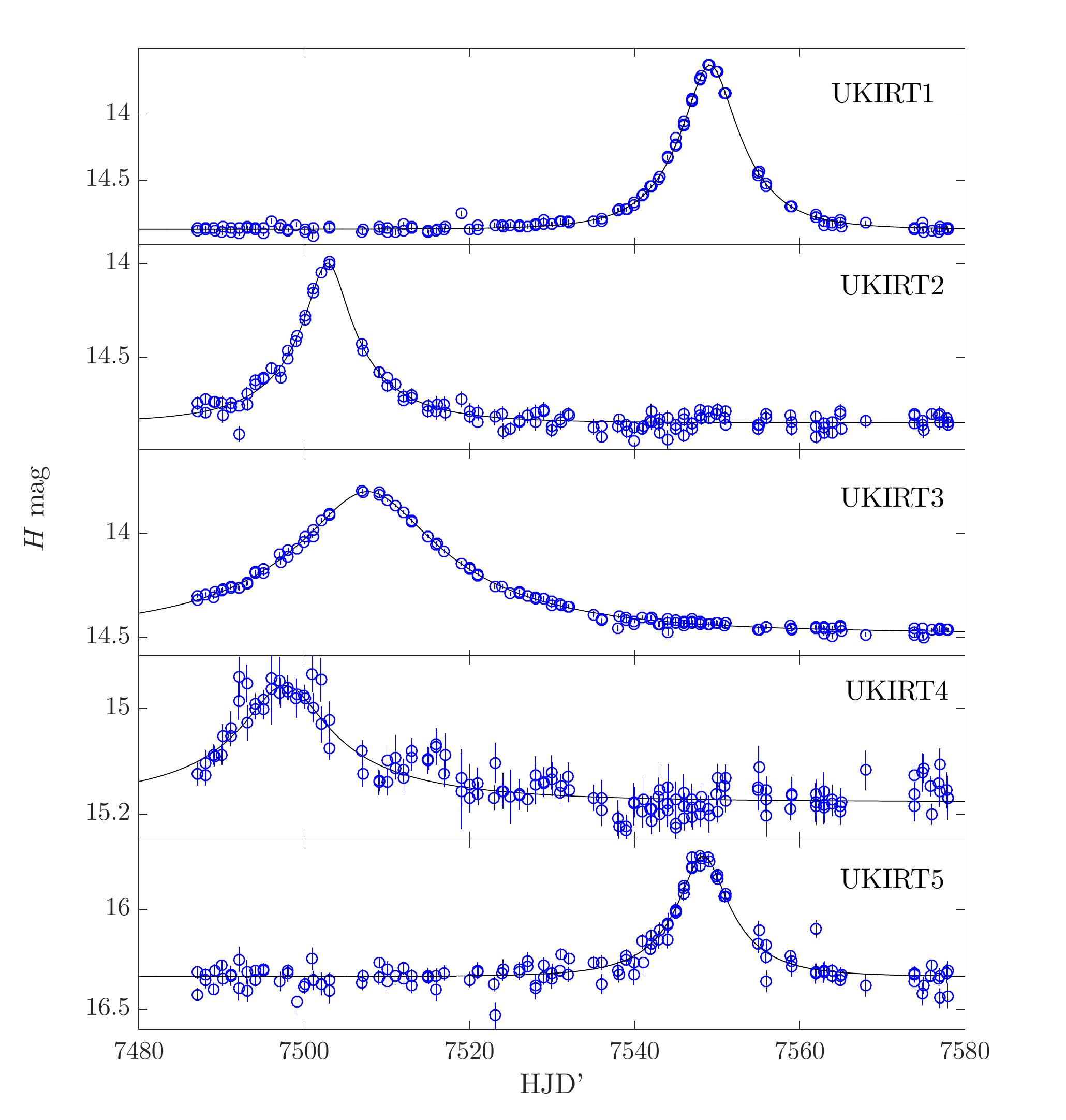}
\caption{Light curves of the five UKIRT microlensing events.
}
\label{fig:lcs}
\end{figure}

\begin{figure}
\plotone{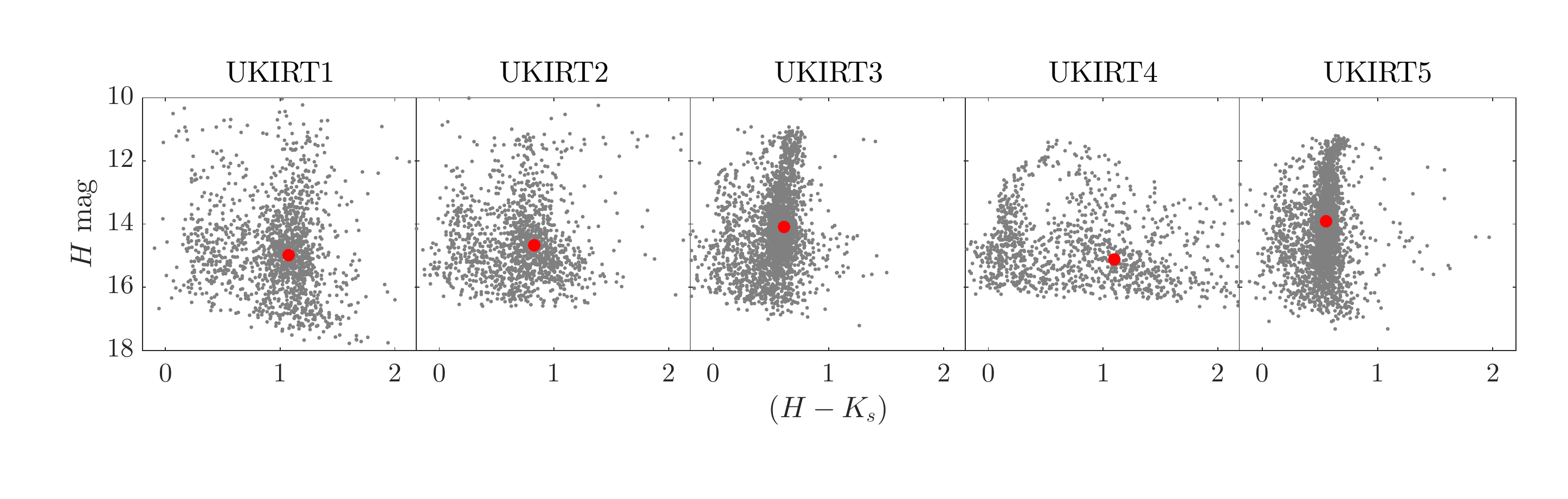}
\caption{VVV color magnitude diagrams (CMD) for the fields of the five events. The red clump position for each is indicated as the red circle.
The CMD for UKIRT4 emphasizes the high degree of differential reddening in this region.
}
\label{fig:cmd}
\end{figure}

\end{document}